%% file: SurfaceEnergy_041220.tex
\documentclass[aps,preprint,prl,superscriptaddress]{revtex4-2} 
\usepackage{amsfonts,amsmath,amssymb}
\usepackage{graphicx}
\usepackage[mathscr]{euscript}

\usepackage{hyperref}
\usepackage{xcolor}
 \hypersetup{
  colorlinks,
  linkcolor={blue},
  citecolor={blue},
  urlcolor={blue}
}

\usepackage{tikz}
\definecolor{lime}{HTML}{A6CE39}
\DeclareRobustCommand{\orcidicon}{%
	\begin{tikzpicture}
	\draw[lime, fill=lime] (0,0) 
	circle [radius=0.16] 
	node[white] {{\fontfamily{qag}\selectfont \tiny ID}};
	\draw[white, fill=white] (-0.0625,0.095) 
	circle [radius=0.007];
	\end{tikzpicture}
	\hspace{-2mm}
}
\foreach \x in {A, ..., Z}{%
 \expandafter\xdef\csname orcid\x\endcsname{\noexpand\href{https://orcid.org/\csname 
 orcidauthor\x\endcsname}{\noexpand\orcidicon}}
}
\newcommand\fr{\frac}
\newcommand\mcA{\mathcal{A}}
\newcommand\mcC{\mathcal{C}}
\newcommand\mcV{\mathcal{V}}
\newcommand\p{\partial}
\newcommand\q{\quad}
\newcommand\sqrb[1]{\sqrt{\smash[b]{#1}}}
\newcommand\ud{\mathrm{d}} 

\begin{document}
\title{Surface Tension and  Energy Conservation in a Moving Fluid}
\author{Tomas Bohr\orcidA{}}\email{tomas.bohr@fysik.dtu.dk}
\affiliation{Department of Physics, Technical University of Denmark, DK-2800 Kgs.\ Lyngby,   
Denmark}
\author{Bernhard Scheichl\orcidB{}}\email{bernhard.scheichl@tuwien.ac.at}
\affiliation{Institute of Fluid Mechanics and Heat Transfer, Technische Universit\"at Wien, 
Getreidemarkt~9, 1060 Vienna, Austria}
\affiliation{AC2T research GmbH, 2700 Wiener Neustadt, Austria}

\date{\today}

\begin{abstract}
The transport of energy in a moving fluid with a simply connected free surface is analyzed, 
taking into account the contribution of surface tension. This is done by following a ``control 
volume" with arbitrary, specified velocity, independent of the flow velocity, and determining 
the rates of energy passing through the boundaries, as well as the energy dissipation in the 
bulk. In particular, a simple conservation equation for the surface area is written down, which 
clearly shows the contribution of the Laplace pressure at the free surface and the tangential 
surface tension forces at its boundary. It emerges as the mechanical conservation law 
for the surface energy in its general form. For a static control volume, all contributions 
from surface tension disappear, except that the pressure has to be modified by the Laplace 
contribution.
\end{abstract}

\maketitle

Capillary effects have been included in flows with free surfaces or interfaces for around 150 years. 
Some famous pioneers are Rayleigh \cite{Ra1878,Ra1879}, Landau and Levich \cite{LaLe42}, 
Taylor \cite{Ta59}, Culick \cite{Cu60}, and Bretherton \cite{Br61} and it is still an active field of research
with outstanding survey papers \cite{Eg97,Boetal09} and the textbook 
\cite{GeBWQu04}. In contrast to flows having free surfaces with objects protruding, where 
surface tension gives rise to tangential forces at the boundaries, or Marangoni flows, where 
gradients of the surface tension contribute to the surface-stress balance, free-surface flows 
include a uniform surface tension only via the discontinuity the normal component of the fluid stresses 
on the interface, commonly referred to as Laplace pressure and 
described, e.g., carefully by Batchelor \cite{Ba00} (\S\,1.9).

Although this is merely a consequence of the governing equations of motion and conventional 
kinematic and dynamic boundary conditions, this procedure has been challenged recently by 
Bhagat {\em et al.}\ \cite{Bhetal18} and Bhagat and Linden \cite{BhLi20}. These authors claim 
that the tangential surface tension contributes to a power term in the integral energy budget 
and should, therefore, be taken into account explicitly. For a {\em static control volume} in 
a steady (stationary) flow, this power is claimed to be
\begin{equation}
 \oint_{\mcC_s}{\bf f}_\gamma\cdot{\bf u}\,\ud\mcC_s.
 \label{fga}
\end{equation} 
Here the closed curve $\mcC_s$ has the infinitesimal arc length $\ud\mcC_s$ and encloses a 
portion of the free surface that forms the control area and confines the control volume,  
${\bf u}$ is the flow velocity on $\mcC_s$ and ${\bf f}_\gamma$ is the external surface tension
force (pr. length).
Indeed, inclusion of such a power term, referring to the external work of surface tension, 
seems intuitively correct and its omission therefore a serious error in all the 
preceding works. However, in this paper we demonstrate that within the continuum hypothesis, where the interface is 
taken as infinitely thin and mass- and inertialess, this is not the case. If the control surface moves with velocity ${\bf U}$, such a term appears
whith ${\bf u}$ is now replaced by ${\bf U}$, so for a static control volume there is no such term. 
Even in the general situation 
of an unsteady flow, the appearance of this term is inconsistent with the advection of surface energy 
combined with the standard work-energy balance for the bulk flow, derived from the momentum 
(Navier--Stokes) equations. This was 
already noticed by Duchesne {\em et al.}\ \cite{DuAnBo19} and Scheichl \cite{Sc20}, but in the present paper we show how this can be seen directly from
from the conservation law for surface area that we derive below.

We consider a free-surface flow of an incompressible fluid (liquid) of uniform density $\rho$,
as sketched in Fig.~\ref{f:1}. 
The free surface forming the interface with the (gaseous) environment is characterised 
by a uniform surface tension $\gamma$. The environment is typically taken as nominally at rest 
and under a constant ambient pressure $p_0$. We assume that the dynamic viscosity $\mu$ for the 
gas is much smaller than that of the liquid, so the
deviatoric stresses from the external gas flow can be neglected (for simplicity).
Let $t$ denote the time and 
${\bf x}=x\,{\bf\hat{x}}+y\,{\bf\hat{y}}+z\,{\bf\hat{z}}=(x,y,z)$ the space vector expressed 
in static Cartesian laboratory coordinates $x$, $y$, $z$ and the associated unit vectors (indicated with hats).
For partial derivatives we use $\nabla=(\p_x,\p_y,\p_z)$ as well as the subscripts $t$, $x$, $y$, $z$,.
We assume gravity to be the only body force at play, with the 
constant gravitational acceleration ${\bf g}=(0,0,-g)$. In the bulk of the flow, 
the Eularian ``material"  velocity field ${\bf u}({\bf x},t)=(u,v,w)$, the 
pressure $p({\bf x},t)$, and the (symmetric) tensor $\bf T$ of the deviatoric 
(viscous) Cauchy stresses, subject to a constitutive relationship, then satisfy the continuity 
{and momentum equations
\begin{align}
 \nabla\cdot{\bf u} &= 0,
 \label{ce}\\
 \rho({\bf u}_t+{\bf u}\cdot\nabla{\bf u}) &= -\nabla p-\rho g\,{\bf\hat{z}}+
 \nabla\cdot{\bf T}.
 \label{me}
\end{align}
For a Newtonian liquid,
${\bf T}=\mu\bigl[\nabla{\bf u}+(\nabla{\bf u})^T\bigr]$ with the dynamic viscosity $\mu$.

\begin{figure}
 \scalebox{0.85}{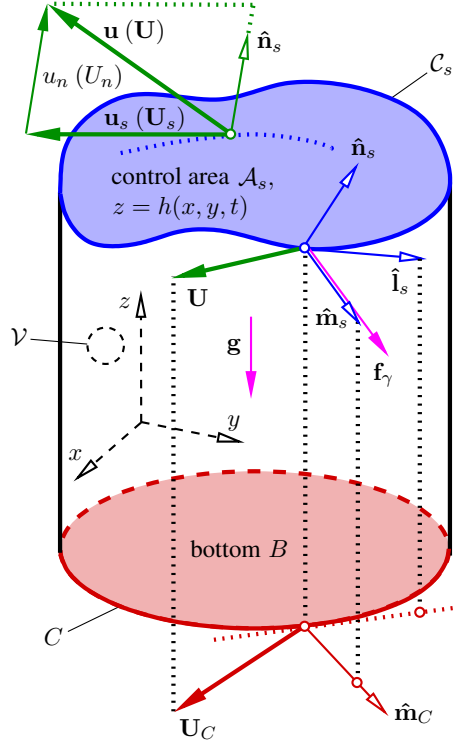}
 \caption{Control volume $\mcV$ with top face $\mcA_s$ lying in free surface, projection 
  $B$ of $\mcA_s$ onto $(x,y)$-plane (bottom); body and surface forces, velocities, unit 
  vectors; dotted: projection and tangent lines.}
 \label{f:1}
\end{figure}

We look at the simply-connected control area $\mcA_s$ of the moving fluid surface with the 
boundary $\mcC_s=\p\mcA_s$ and consider also their normal projections $B$ and $C=\p B$ onto 
on the $(x,y)$-plane (the ``bottom''). The free surface is described
simplistically as a function $z=h(x,y,t)$. The value of ${\bf u}$ on the surface can be 
decomposed into a component {\em in} the surface (i.e., in the tangent plane of the 
instantaneous surface) ${\bf u}_s(x,y,t)$ and a normal component $u_n(x,y,t)$ along the outward 
normal direction, $n_s$, and surface unit normal
\begin{equation}
 {\bf\hat{n}}_s(x,y,t)\!=\!\frac{(-h_x,-h_y,1)}{\sqrb{1+|\nabla h|^2}}\!=\!
 \frac{{\bf\hat{z}}-{\nabla h}}{\sqrb{1+|\nabla h|^2}}.
 \label{ns}
\end{equation}
We furthermore introduce the vectors
\begin{equation}
 {\bf\hat{l}}_s =\ud{\bf x}/\ud\mcC_s ,\q 
 {\bf\hat{m}}_s ={\bf\hat{l}}_s\times{\bf\hat{n}}_s \q
 {\bf f}_\gamma = \gamma \,  {\bf\hat{m}}_s
 \label{lms}
\end{equation} 
where $\ud\mcC_s$ is the infinitesimal arc length on $\mcC_s$, ${\bf\hat{l}}_s$ a unit tangent vectior on $\mcC_s$, $ {\bf\hat{m}}_s$ is
a unit vector perpendicular to $\mcC_s$, and  ${\bf f}_\gamma $ is the tangential capillary force pr. length from the surface region outside  $\mcC_s$.
Similarly, in the $(x,y)$-plane, with
infinitesimal arc length $\ud C$ of $C$, its outer normal ${\bf\hat{m}}_C$, and the projection 
of ${\bf\hat{n}}_s$ are
\begin{equation}
 {\bf\hat{m}}_C=\fr{(\ud y,-\ud x,0)}{\ud C},\q 
 {\bf n}_b=\fr{-\nabla h}{\sqrb{1+|\nabla h|^2}}.
 \label{mnb}
\end{equation}
Further $\bf u$ satisfies the  kinematic free-surface boundary condition
\begin{equation}
 h_t=w-u h_x-v h_y=u_n\sqrt{1+|\nabla h|^2}
 \label{kbc}
\end{equation}
on $\mcA_s$, so $u_n$ is only nonzero when $h$ is explicitly time-dependent. The
dynamic boundary condition expresses continuity of the total stress across $\mcA_s$  in the form
(see, e.g., \cite{LaLi87}, \S\,61)
\begin{equation}
 (p-p_0)\,{\bf{\hat{n}}_s}-{\bf T}\cdot{\bf{\hat{n}}_s}=  \Delta p_L\,{\bf{\hat{n}}_s},
 \label{pl}
\end{equation}
where $\Delta p_L$ is the Laplace pressure involving the double local mean curvature 
$\kappa$ of $\mcA_s$ as
\begin{equation}
 \Delta p_L= \gamma  \kappa= \gamma \nabla\cdot{\bf \hat{n}}_s= \gamma \nabla\cdot{\bf n}_b
 \label{ka}
\end{equation}
(where the sign of $\kappa$ is chose so that $\kappa>0$ and $\nabla^2 h<0$ for a convex, droplet-shaped 
surface). The equivalence of the local stress equilibrium \eqref{pl} on $\mcA_s$ with its 
global counterpart 
\begin{equation}
 \int_{\mcA_s} \bigl[(p-p_0)\,{\bf{\hat{n}}_s}-{\bf T}\cdot{\bf{\hat{n}}_s}\bigr]\,\ud A_s=
 -\!\oint_{\mcC_s} {\bf f}_\gamma\,\ud\mcC_s
 \label{plg}
\end{equation}
is obtained using Stokes' theorem in the form
\begin{equation}
\label{Stokes}
\!\oint_{\mcC_s} {\bf\hat{m}}_s\,\ud\mcC_s = \!\oint_{\mcC_s} {\bf\hat{l}}_s\times{\bf\hat{n}}_s\,\ud\mcC_s = -  \int_{\mcA_s} (\nabla \cdot {\bf\hat{n}}_s) {\bf\hat{n}}_s\,\ud A_s
\end{equation}
as shown, e.g., in \cite{Po11}, with the counterintuitive result that an integral over 
tangential vectors points in the direction of the normals.

It is our goal to combine the (hitherto unappreciated) conservation law for the 
surface energy contained in $\mcA_s$ with that for the mechanical energy of the bulk flow in 
an arbitrarily moved, simply-connected control volume $\mcV$ bounded by $\mcA_s$ at the free surface. 
We denote the velocity field moving the closed boundary $\p\mcV$ as
${\bf U}({\bf x},t)=(U,V,W)$ and denote its components tangential and normal to $\mcA_s$  as $U_s$ and $U_n$, respectively. 
Analogously,  $C$ moves with ${\bf U}_C$, the projection of $\bf U$ evaluated on $\mcC_s$ on the $(x,y)$-plane.
Importantly, $\bf U$ must be kinematically compatible to the material flow field 
$\bf u$ in the sense that any point on the free surface considered to move with $\bf U$ 
remains in the surface for all times. That is, $\bf U$ satisfies
\begin{equation}
 h_t=W-U h_x-V h_y=U_n\sqrt{1+|\nabla h|^2}
 \label{kbcu}
\end{equation}
on $\mcA_s$, in analogy to \eqref{kbc}. Thus a completely static control volume $\mcV$ is of course only possible for a steady flow.

The surface energy stored in $\mcA_s$ is simply $E_s=\gamma A_s$ where
\begin{equation}
 A_s=\int_{\mcA_s}\,\ud A_s=\int_B\sqrt{1+\left|\nabla h\right|^2}\,\ud x\ud y
 \label{as}
\end{equation}
is the time-dependent area of $\mcA_s$. The integral conservation law for the surface 
energy is found by analysing $\gamma\,\ud A_s/\ud t$ in terms of $\bf U$, which 
transports $\mcV$ and thus $\mcA_s$. Using Leibnitz' integral rule, the time derivative of 
\eqref{as} is
\begin{align}
 \fr{\ud A_s}{\ud t} =& \int_B \p_t\sqrt{1+|\nabla h|^2}\,\ud x\ud y \nonumber\\[2pt] 
 +& \oint_C \sqrt{1+|\nabla h|^2}\,{\bf U}_C\cdot{\bf\hat{m}}_C\,\ud C.
 \label{astb}
\end{align}
The first term to the right can be rewritten as
\begin{widetext}
\begin{align}
 \int_B \fr{\nabla h_t\cdot\nabla h}{\sqrb{1+|\nabla h|^2}}\,\ud x\ud y &=
 \int_B \left[-h_t\nabla\cdot\left(\fr{\nabla h}{\sqrb{1+|\nabla h|^2}}\right)+
 \nabla\cdot\left(\fr{h_t\nabla h}{\sqrb{1+|\nabla h|2}}\right)\right]\ud x\ud y
 \nonumber\\[2pt]
 &= \int_B h_t\,\kappa\,\ud x\ud y+
 \oint_C \fr{h_t}{\sqrb{1+|\nabla h|^2}}\nabla h\cdot{\bf\hat{m}}_C\,\ud C.
 \label{astb1}
\end{align}
\end{widetext}
The last equality follows from \eqref{ka} and the divergence theorem. All in all, we have
\begin{align}
 \fr{\ud A_s}{\ud t} &= \int_B h_t\,\kappa\,\ud x\ud y \nonumber\\[2pt]
 &+ \oint_C \left(\sqrt{1+|\nabla h|^2}\,{\bf U}+
    \fr{h_t}{\sqrb{1+|\nabla h|^2}}\nabla h\right)\cdot{\bf\hat{m}}_C\,\ud C.
 \label{astbf}
\end{align}
The expressions \eqref{lms} combined with \eqref{ns} yield the explicit representation
\begin{equation}
 {\bf\hat{m}}_s=\fr{\left[(1+h_y^2)\ud y,-(1+h_x^2)\ud x,h_x\ud y-h_y\ud x)\right]}
 {\sqrb{1+|\nabla h|^2}\,\ud\mcC_s}.
 \label{msdcs}
\end{equation} 
and together with the kinematic boundary condition \eqref{kbcu} this leads to the simple result 
\begin{equation}
 \fr{\ud A_s}{\ud t}=\int_{\mcA_s} U_n\,\kappa\,\ud A_s+
 \oint_{\mcC_s} {\bf U}\cdot{\bf\hat{m}}_s\,\ud\mcC_s.
 \label{ast}
\end{equation}
Actually this result could have been obtained more directly and intuitively by considering independently the three infinitesimally small 
temporal (linear) variations of the shape of the material surface element $\ud A$: 1. dilatation 
normal to $\mcA_s$ (inflation), 2. distortion in its tangential directions (stretching and 
straining) and 3.  tangential solid-body rotation. Since the latter has no effect, this leaves 
us with the two contributions on the right side of \eqref{ast}, where the first term 
originates in the relation $(\delta\ud A_s/\delta n_s)/\ud A_s$ (see, e.g., 
\cite{ChLuNi06} and \cite{LaLi87}, \S\,61) and the last term describes the tangential change of 
$A$ as the change of area adjacent to $\mcC_s$.

It is instructive to write down the analog of \eqref{ast} for a planar flow where the free 
surface is a line $z=h(x,t)$ as shown Fig.~\ref{f:2}.
The control volume (area) has  endpoints $x=x_1(t)$ and $x=x_2(t)>x_1$ on the free surface move with the imposed velocity $U(x,t)$.
The analog of the surface are, i.e. the length of the free surface segment is
\begin{equation}
 L_s=\int_{\mathcal{L}_s} \ud L_s=\int_{x_1}^{x_2} \sqrt{1+h_x^2}\,\ud x
 \label{ls}
\end{equation} 
and the curvature $\kappa_1$ is
\begin{equation}
 \kappa_1=-\fr{\p}{\p x}\left(\fr{h_x}{\sqrb{1+h_x^2}}\right)= 
 -\fr{h_{xx}}{(1+ h_x^2)^{3/2}}.
 \label{ka1}
\end{equation}
If $d$ denotes its depth in the $y$-direction, 
$E_s=\gamma L_s d$ is the corresponding surface energy. In order to consider the advection of 
a point on this segment by a velocity field ${\bf U}(x,y,t)=(U,0,W)$, we take its 
$x$-coordinate as time-dependent, thus its $z$-coordinate as $h(x(t),t)$. On $\mcC_s$, 
$U=\dot x$ and $W=h_t+\dot x h_x$ by \eqref{kbcu}. It is evident that 
${\bf U}\cdot{\bf\hat{m}}_s$ vanishes on the points of $\mcC_s$ with $x$-coordinates located 
between $x_1$ and $x_2$. Here
${\bf{\hat m}}_s={(\bf\hat{x}}+h_x\,{\bf\hat{z}})/\sqrb{1+h_x^2}$ and $\bf U$ are collinear if 
$h_t$ vanishes. These results help us to rewrite \eqref{ast} in this case as
\begin{align}
\nonumber
 \fr{\ud L_s}{\ud t} &= \int_{x_1}^{x_2} h_t\,\kappa_1\,\ud x + \left[\sqrt{1+h_x^2}\,\dot x+\fr{h_t\,h_x}{\sqrb{1+h_x^2}}\right]_{x=x_1}^{x=x_2}\\
 \label{astpf}
 & =\int_{\mathcal{L}_s} U_n\,\kappa_1\,\ud L_s+
\left[{\bf U}\cdot{\bf\hat{m}}_s\right]_{x=x_1}^{x=x_2}
\end{align}
in direct analogy with \eqref{astbf} by
noticing the projection ${\bf U}_C=\dot x_{1,2}\,{\bf\hat{x}}$ of $\bf U$ onto the endpoints. 
Alternatively, one obtains \eqref{astpf} from \eqref{ast} in straightforward manner under 
the assumptions of a planar flow and $\mcC$ choosen such that $C$ is a rectangle with a side of 
length $x_2-x_1$ parallel to the $x$-axis.

\begin{figure}
 \scalebox{0.75}{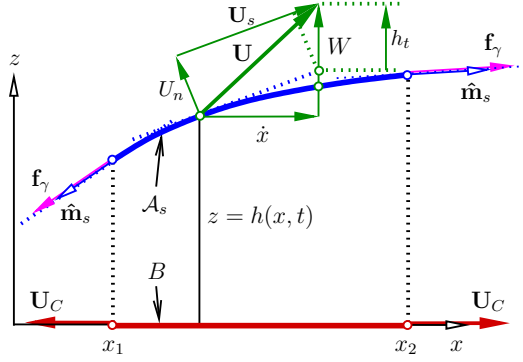}
 \caption{Planar flow: projection onto $(x,z)$-plane; for legend see Fig.\,\ref{f:1}.}
 \label{f:2}
\end{figure}

We next remark some important consequences of \eqref{ast}. First we derive the differential 
version using the divergence theorem. With the normal projection 
$\nabla_s=\nabla-{\bf\hat{n}}_s\p_n$ of $\nabla$ onto $\mcA_s$ acting on the tangential flow 
field ${\bf U}_s={\bf U}-U_n{\bf\hat{n}}_s$,
\begin{equation}
 \oint_{\mcC_s} {\bf U}\cdot{\bf\hat{m}}_s\,\ud\mcC_s=
 \int_{\mcA_s} \nabla_s\cdot{\bf U}_s\,\ud A_s.
 \label{dts}
\end{equation}
Using $\p_n\equiv{\bf\hat{n}}_s\cdot\nabla$, we expand the divergence in the surface 
integral in \eqref{dts} as
$\nabla_s\cdot{\bf U}_s= \nabla\cdot{\bf U}-{\bf\hat{n}}_s\cdot~({\bf\hat{n}}_s\cdot\nabla{\bf U})-\nabla_s\cdot(U_n{\bf\hat{n}}_s)$. Since $\nabla_s U_n$ lies in the surface and 
$\nabla_s\cdot{\bf\hat{n}}_s\equiv\nabla\cdot{\bf\hat{n}}_s=\kappa$, cf.\ \eqref{ka}, 
\eqref{ast} takes on the form
\begin{equation}
 \fr{\ud A_s}{\ud t}=\int_{\mcA_s} [U_n\nabla\cdot{\bf\hat{n}}_s+\nabla_s\cdot{\bf U}_s]\,\ud A_s= \int_{\mcA_s} 
 \bigl[\nabla\cdot{\bf U}-{\bf\hat{n}}_s\cdot\left(\nabla{\bf U}\right)\cdot {\bf\hat{n}}_s\bigr]\,\ud A_s.
 \label{asta}
\end{equation}
As $\mcA_s$ is chosen arbitrarily, the relative temporal rate of change of an infinitesimal 
area element must be
\begin{equation}
 (\delta\ud A_s/\delta t)/\ud A_s= U_n\nabla\cdot{\bf\hat{n}}_s+\nabla_s\cdot{\bf U}_s=
 \nabla\cdot{\bf U}-{\bf\hat{n}}_s\cdot\left(\nabla{\bf U}\right)\cdot {\bf\hat{n}}_s
 \label{astl}
\end{equation}
For $\bf U$ identical to the material flow field $\bf u$, this local form of \eqref{asta} 
was already derived by Batchelor \cite{Ba00}, eqs.~(3.1.5)--(3.1.8), without taking notice of 
its integral formulation. For an incompressible material flow field, it takes the form
$(\delta\ud A_s/\delta t)/\ud A_s=
 -{\bf\hat{n}}_s\cdot\left(\nabla{\bf u}\right)\cdot {\bf\hat{n}}_s$
and these local forms makes explicit the required invariance under the Galilean 
boost, i.e., for (\ref{astl}): ${\bf U}\mapsto{\bf U}+{\bf C}$, ${\bf x}\mapsto{\bf x}+{\bf 
C}t$, which is evident  neither from \eqref{ast} nor from \eqref{astbf}.

Finally, for a stationary velocity field $\bf U$, the curvature term in \eqref{ast} is zero. 
Interestingly and counterintuitively, then $\ud A_s/\ud t$ does not necessarily vanish even if 
${\bf U\equiv \bf u}$ represents an incompressible flow satisfying \eqref{ce}. This suppresses 
the dilatation of the fluid particles but not their stretching and straining {\em in} the 
surface governed by the surface divergence in \eqref{dts}. Therefore, the streamlines lying in 
the surface then can diverge/converge, which explains the in general non-vanishing 
$\ud A_s/\ud t$. 

In the remainder of this letter, we discuss the dynamic implications of the kinematic 
relationship \eqref{ast}. Multiplying \eqref{ast} with $\gamma$ and using the expressions for 
the Laplace pressure in \eqref{pl} and the vectorial surface tension, 
${\bf f}_\gamma=\gamma\,{\bf\hat{m}}_C$, results in the integral transport equation for 
the surface energy convected by $\bf U$:
\begin{equation}
 \fr{\ud E_s}{\ud t}=\int_{\mcA_s} \Delta p_L U_n\,\ud A_s+
 \oint_{\mcC_s} {\bf f}_\gamma\cdot{\bf U}\,\ud\mcC_s.
 \label{est}
\end{equation}
This has an appealing and convincing physical interpretation as a law of mechanical energy 
conversion for the piece $\mcA_s$ of the inertialess free fluid surface: the work done by the 
Laplace pressure on $\mcA_s$ and by the tangential surface tension forces at its boundary 
$\mcC_s$, on the surface moving with the imposed velocity field ${\bf U}$, changes the 
potential internal energy, i.e.\ the surface energy $E_s$, contained in $\mcA_s$. Therefore, 
\eqref{est} represents the precise justification for the Laplace pressure and the existence of 
the tangential surface tension forces. It is readily seen that \eqref{est} can also be 
derived from the pointwise relationships \eqref{ka}, after multiplication with $U_n$, and 
\eqref{astl}. Note that the inclusion of Marangoni effect (inhomogeneous $\gamma$) would lead 
to an extension of \eqref{est} incorporate the transport of $\gamma$ as a non-uniform passive 
scalar \cite{St90}.

The conservation law \eqref{est} is the basis for the subsequent careful re-examination of 
the work-energy balance for the entire volume $\mcV$  {\em including} its top face $\mcA_s$ once ${\bf U}$ is 
specified appropriately.  To do this, we first briefly recall the 
derivation of the classical work-energy theorem as stated, e.g., in 
eq.\ (16.2) in \cite{LaLi87}. More precisely, it denotes the transport of the overall 
mechanical energy of the bulk flow,
\begin{equation}
 E_m=\int_\mcV e_m\,\ud V,\q e_m=\fr{\rho|{\bf u}|^2}{2}+\rho g z,
 \label{em}
\end{equation}
where the energy density $e_m$ consists of the kinetic and the potential energy densities. 
Taking the inner product of ${\bf u}$ and the momentum equation  \eqref{me} together with
the incompressibility condition \eqref{ce} and the identity $w\equiv\nabla\cdot(z{\bf u})$
leads to the differential form of the energy equation: 
\begin{equation}
 \p_t e_m=\nabla\cdot[{\bf T}\cdot{\bf u}-(p+e_m){\bf u}]-{\bf T}\cdot\nabla{\bf u}.
 \label{eeqd}
\end{equation}
Integration of \eqref{eeqd} over $\mcV$ and applying the divergence theorem gives
\begin{equation}
 \int_\mcV \p_t e_m\,\ud V=
 \oint_{\p\mcV} [{\bf T}\cdot{\bf u}-(p+e_m){\bf u}]\cdot{\bf\hat{n}}\,\ud A-\!
 \int_\mcV {\bf T}\cdot\nabla{\bf u}\,\ud V
 \label{eeq0}
\end{equation}
where the first term represents the advected energy, and the last term represents 
the (non-negative) integral dissipation per time unit due to the deviatoric stresses, commonly 
denoted with $\dot\Phi$. We have also introduced the infinitesimal area element $\ud A$ of 
$\p\mcV$ and its outward unit normal $\bf\hat{n}$. We note that, according to the 
derivation of \eqref{eeq0}, the stresses acting on $\p\mcV$ are taken as their limiting values 
approached from inside $\mcV$. For the bulk flow this makes no difference due to continuity,
but when we include the free surface it becomes important. The transport equation then follows from the classical 
Leibnitz--Reynolds transport theorem (cf.\ \cite{Le07}, eqs.\ (3.1.3)--(3.1.5) in \cite{Ba00}):
\begin{align}
\nonumber
 \fr{\ud E_m}{\ud t}&=\int_\mcV \p_t e_m\,\ud V+
 \oint_{\p\mcV} e_m{\bf U}\cdot{\bf\hat{n}}\,\ud A\\
 &=\oint_{\p\mcV} 
 \bigl[e_m({\bf U}-{\bf u})+{\bf T}\cdot{\bf u}-p\,{\bf u}\bigr] \cdot{\bf\hat{n}}\,\ud A- \dot\Phi
\label{eeq}
\end{align}
This work-energy principle states that the temporal change of $E_m$ is 
compensated by the flux of mechanical energy into $\mcV$ and the external work done by the 
total stresses acting on $\p\mcV$, due to the material velocity ${\bf u}$, and the associated 
internal work of the deviatoric (viscous) stresses, equal to the negative mechanical 
dissipation.  It equals zero for a stationary flow. 

For many application, especially for stationary flows, it is valuable 
to consider a cylindrical control volume $\mcV$ aligned with $z$-direction and having the 
patch $\mcA_s$ of the free surface as its top boundary, as indicated in Fig.~\ref{f:1}. In the 
general, non-stationary situation its lateral sides facing the bulk of the flow and $\mcA_s$ 
are both moving. In this important case of a static control volume, its projection $B$ and thus 
$C$ stay stationary while $\mcA_s$ is moving though with the normal component of the material flow
(when the flow is not stationary).
Thus ${\bf U}={\bf 0}$ on the lateral sides of $\mcV$ 
and  ${\bf U}_C\equiv{\bf 0}$ but ${\bf U}=u_n{\bf\hat{n}}_s$, ${\bf\hat{n}}={\bf\hat{n}}_s$, 
$\ud A=\ud A_s$ on its top face $\mcA_s$. In this case, 
\eqref{eeq} is written as
\begin{align}
\nonumber
 &\fr{\ud E_m}{\ud t}=\fr{\ud}{\ud t}\int_0^{h(x,y,t)} \ud z \int_B e_m\,\ud x\ud y \\
 &= -\!\int_{\mcA_s} \Delta p_L  u_n\,\ud A_s+\!\int_{\p\mcV_r}  
 \left[{\bf T}\cdot{\bf u}-(e_m+p-p_0){\bf u}\right]\cdot{\bf\hat{n}}\,\ud A -\dot\Phi
 \label{eeqs}
\end{align}
where $\p\mcV_r$ denotes the remaining boundary, i.e., the boundary excluding the free surface, 
and where we have subtracted the constant $p_0$ from the pressure everywhere since 
$\oint_{\p\mcV} {\bf u}\cdot{\bf\hat{n}}\,\ud A=0$ by continuity and incompressibility.To 
determine the total energy budget, we note that the Laplace pressure acts on both sides of 
the interface separating $\mcA_s$ and $\mcV$. But the powers it exerts on $\mcA_s$, see 
\eqref{est}, and on $\mcV$, see \eqref{eeqs},  differ in theirs signs and therefore cancel. 
Since we saw from the comparison with \eqref{astb} that the loop integral in \eqref{ast} 
vanishes for ${\bf U}_C\equiv{\bf 0}$, this yields the total energy 
budget in the form
\begin{equation}
 \fr{\ud(E_m+E_s)}{\ud t}= \!\int_{\p\mcV_r} \left({\bf T}\cdot{\bf u}-(e_m+p-p_0){\bf u}\right)\cdot{\bf\hat{n}}\,\ud A-\dot\Phi.
 \label{eeqt}
\end{equation}
For stationary flow $\ud E_s/\ud t$ vanishes as does $u_n$; see \eqref{ast}. Then the 
energy equation balance \eqref{eeqt}, combining that of the bulk and the surface flow, becomes 
indistinguishable from that just governing the bulk flow, \eqref{eeqs}. Correspondingly, the 
only trace of capillarity then comes from the pressure $p$, which must include the Laplace pressure to respect the boundary condition
(\ref{pl})-(\ref{ka}). This is the standard approach, and it validity was recently emphasized in \cite{DuAnBo19} and \cite{Sc20}.
If this seems strange, one should 
keep in mind that surface area is not advected like mass or kinetic energy. In the bulk of the 
flow, each fluid particle carries a particular mass and kinetic energy with it, but it does 
{\em not} carry a particular surface area with it. 

The additional energy term \eqref{fga} for a static 
control volume in a steady flow was originally postulated by \cite{Bhetal18} without explanation.
In their follow-up paper \cite{BhLi20} they argue that it actually stems from the viscous 
advection terms in \eqref{eeq}.
Writing the global stress balance \eqref{plg} in the form (their eq.~(3.4) in our notation)
\begin{equation}
\int_{\mcA_s} (p - p_0) {\bf\hat{n}}_s \ud A_s - \int_{\mcA_s} {\bf\hat{n}}_s\cdot{\bf T}\,\ud 
A_s + \gamma \!\oint_{\mcC_s} {\bf \ud l} \times {\bf\hat{n}}_s=0,
 \label{bl3.4}
\end{equation}
they then take the inner product with $\bf u$ {\em inside} both integrals to get
\begin{equation}
\int_{\mcA_s} (p - p_0) {\bf u}\cdot{\bf\hat{n}}_s \ud A_s - \int_{\mcA_s} {\bf u}\cdot{\bf T}\cdot {\bf\hat{n}}_s\,\ud A_s + \gamma \!\oint_{\mcC_s} {\bf u}\cdot({\bf \ud l} \times {\bf\hat{n}}_s)=0
 \label{bl3.4b}
\end{equation}
and conclude, since ${\bf u}\cdot{\bf\hat{n}}_s $ (for a stationary flow), that 
\begin{equation}
 \int_{\mcA_s} {\bf u}\cdot{\bf T}\cdot {\bf\hat{n}}_s\,\ud A_s =- \gamma \!\oint_{\mcC_s} {\bf 
u}\cdot({\bf \ud l} \times {\bf\hat{n}}_s),
 \label{bl3.4c}
\end{equation}
relating viscous and capillary effects and reinstating the viscous term that was dropped from 
the free surface part of the energy equation \eqref{eeqt}.
However, multiplying by ${\bf u}$ inside the integral, which might seem a legal operation since the area of integration is arbitrary, is actually {\em illegal}
since one of the integrals is a loop integral. Rewriting the loop integral in \eqref{bl3.4} as 
a surface integral, using the identity \eqref{Stokes}, we get (as in 
\eqref{plg}--\eqref{Stokes})
 \begin{equation}
\int_{\mcA_s} \big[(p - p_0) {\bf\hat{n}}_s -  {\bf T}\cdot {\bf\hat{n}}_s- \gamma  (\nabla \cdot {\bf\hat{n}}_s) {\bf\hat{n}}_s \big]\ud A_s=0
 \label{bl3.4d}
\end{equation}
and we can now take ${\mcA_s}$ as any shape we want and conclude -- since the first and last terms are explicitly orthogonal to ${\bf u}$ --  that 
\begin{equation}
 \int_{\mcA_s} {\bf u}\cdot{\bf T}\cdot{\bf\hat{n}}_s\,\ud A_s  = 0
 \label{bl3.4e}
\end{equation}
Indeed the integrand ${\bf u}_s\cdot{\bf T}\cdot{\bf\hat{n}}_s$ (which we removed from the free 
surface to get \eqref{eeqt}) must be zero everywhere on the interface, which is simply the 
shear part of the usual dynamic boundary condition \eqref{pl} on a free surface, and does {\em 
not} involve the surface tension. 

As we mentioned above, it is perhaps strange that the surface tension forces ${\bf f}_\gamma$ 
in \eqref{est} do no work when the control volume does not move, since these forces are still 
acting on a moving fluid. 
The reason for this is that the surface energy, being proportional to the surface area 
\eqref{as}, does {\em not} depend on the fluid velocity ${\bf u}$. In other words, the 
surface energy has no kinetic energy term, which is due to the fact that the surface, in our 
idealized continuum approach, is infinitely thin and thus has no mass. But how can it then have 
an energy? The reason for this is found in the fact that the surface energy, coming from the 
breaking of bonds between the liquid molecules, is at least of the order of $kT$ pr.\ particle 
or $RT$ pr.\ mol, i.e. around 2\,kJ/mol ($k$, $R$, and $T$ denote the Boltzmann and the 
universal gas constant, and the absolute temperature, respectively), whereas the kinetic 
energy of matter, say water, moving at 1 m/s is around 0.02\,J/mol. The ratio of these two 
energies of around 10$^5$ allows us to neglect the kinetic energy of the organized
motion represented by the velocity field ${\bf u}$ compared to the surface energy for the same 
amount of matter -- as opposed to the kinetic energy of the random thermal motion, which is 
also of the order of $kT$ pr.\ particle.

\bigskip
This work was co-funded by the project COMET {\em InTribology}, FFG-No.~872176 (project 
coordinator: AC2T research GmbH, Austria). Tomas Bohr is grateful to Anders Andersen and 
Alexis Duchesne for many valuable discussions and to Jens Eggers for useful comments.

The authors contributed equally to this work.

\end{document}

%% file: conf.pdf_t
\begin{picture}(0,0)%
\includegraphics{conf.pdf}%
\end{picture}%
\setlength{\unitlength}{4144sp}%
\begingroup\makeatletter\ifx\SetFigFont\undefined%
\gdef\SetFigFont#1#2#3#4#5{%
  \reset@font\fontsize{#1}{#2pt}%
  \fontfamily{#3}\fontseries{#4}\fontshape{#5}%
  \selectfont}%
\fi\endgroup%
\begin{picture}(3243,5252)(788,-5576)
\put(1991,-5512){\makebox(0,0)[lb]{\smash{{\SetFigFont{11}{13.2}{\rmdefault}{\mddefault}{\updefault}{\color[rgb]{0,0,0}${\bf U}_C$}%
}}}}
\put(3504,-5400){\makebox(0,0)[lb]{\smash{{\SetFigFont{11}{13.2}{\rmdefault}{\mddefault}{\updefault}{\color[rgb]{0,0,0}${\bf\hat{m}}_C$}%
}}}}
\put(3198,-1421){\makebox(0,0)[lb]{\smash{{\SetFigFont{11}{13.2}{\rmdefault}{\mddefault}{\updefault}{\color[rgb]{0,0,0}${\bf\hat{n}}_s$}%
}}}}
\put(1039,-4875){\makebox(0,0)[lb]{\smash{{\SetFigFont{11}{13.2}{\rmdefault}{\mddefault}{\updefault}{\color[rgb]{0,0,0}$C$}%
}}}}
\put(2331,-3335){\makebox(0,0)[lb]{\smash{{\SetFigFont{11}{13.2}{\rmdefault}{\mddefault}{\updefault}{\color[rgb]{0,0,0}$y$}%
}}}}
\put(1213,-3574){\makebox(0,0)[lb]{\smash{{\SetFigFont{11}{13.2}{\rmdefault}{\mddefault}{\updefault}{\color[rgb]{0,0,0}$x$}%
}}}}
\put(803,-2765){\makebox(0,0)[lb]{\smash{{\SetFigFont{11}{13.2}{\rmdefault}{\mddefault}{\updefault}{\color[rgb]{0,0,0}$\mcV$}%
}}}}
\put(1564,-2502){\makebox(0,0)[lb]{\smash{{\SetFigFont{11}{13.2}{\rmdefault}{\mddefault}{\updefault}{\color[rgb]{0,0,0}$z$}%
}}}}
\put(2538,-672){\makebox(0,0)[lb]{\smash{{\SetFigFont{11}{13.2}{\rmdefault}{\mddefault}{\updefault}{\color[rgb]{0,0,0}${\bf\hat{n}}_s$}%
}}}}
\put(3759,-834){\makebox(0,0)[lb]{\smash{{\SetFigFont{11}{13.2}{\rmdefault}{\mddefault}{\updefault}{\color[rgb]{0,0,0}$\mcC_s$}%
}}}}
\put(2619,-4256){\makebox(0,0)[lb]{\smash{{\SetFigFont{11}{13.2}{\rmdefault}{\mddefault}{\updefault}{\color[rgb]{0,0,0}$B$}%
}}}}
\put(2044,-2484){\makebox(0,0)[lb]{\smash{{\SetFigFont{11}{13.2}{\rmdefault}{\mddefault}{\updefault}{\color[rgb]{0,0,0}${\bf U}$}%
}}}}
\put(1029,-919){\makebox(0,0)[lb]{\smash{{\SetFigFont{11}{13.2}{\rmdefault}{\mddefault}{\updefault}{\color[rgb]{0,0,0}$u_n\:(U_n)$}%
}}}}
\put(2947,-2619){\makebox(0,0)[lb]{\smash{{\SetFigFont{11}{13.2}{\rmdefault}{\mddefault}{\updefault}{\color[rgb]{0,0,0}${\bf\hat{m}}_s$}%
}}}}
\put(3485,-2379){\makebox(0,0)[lb]{\smash{{\SetFigFont{11}{13.2}{\rmdefault}{\mddefault}{\updefault}{\color[rgb]{0,0,0}${\bf\hat{l}}_s$}%
}}}}
\put(3355,-3006){\makebox(0,0)[lb]{\smash{{\SetFigFont{11}{13.2}{\rmdefault}{\mddefault}{\updefault}{\color[rgb]{0,0,0}${\bf f}_\gamma$}%
}}}}
\put(2327,-2795){\makebox(0,0)[lb]{\smash{{\SetFigFont{11}{13.2}{\rmdefault}{\mddefault}{\updefault}{\color[rgb]{0,0,0}$\bf g$}%
}}}}
\put(1456,-598){\makebox(0,0)[lb]{\smash{{\SetFigFont{11}{13.2}{\rmdefault}{\mddefault}{\updefault}{\color[rgb]{0,0,0}${\bf u}\:(\bf U)$}%
}}}}
\put(1461,-1214){\makebox(0,0)[lb]{\smash{{\SetFigFont{11}{13.2}{\rmdefault}{\mddefault}{\updefault}{\color[rgb]{0,0,0}${\bf u}_s\:({\bf U}_s)$}%
}}}}
\put(1521,-1838){\makebox(0,0)[lb]{\smash{{\SetFigFont{11}{13.2}{\rmdefault}{\mddefault}{\updefault}{\color[rgb]{0,0,0}$z=h(x,y,t)$}%
}}}}
\put(2406,-1636){\makebox(0,0)[lb]{\smash{{\SetFigFont{11}{13.2}{\rmdefault}{\mddefault}{\updefault}{\color[rgb]{0,0,0}$\mcA_s$,}%
}}}}
\end{picture}%

%% file: plan.pdf_t
\begin{picture}(0,0)%
\includegraphics{plan.pdf}%
\end{picture}%
\setlength{\unitlength}{4144sp}%
\begingroup\makeatletter\ifx\SetFigFont\undefined%
\gdef\SetFigFont#1#2#3#4#5{%
  \reset@font\fontsize{#1}{#2pt}%
  \fontfamily{#3}\fontseries{#4}\fontshape{#5}%
  \selectfont}%
\fi\endgroup%
\begin{picture}(4207,2878)(1211,-7625)
\put(2809,-6539){\makebox(0,0)[lb]{\smash{{\SetFigFont{12}{14.4}{\rmdefault}{\mddefault}{\updefault}{\color[rgb]{0,0,0}$z=h(x,t)$}%
}}}}
\put(1636,-6510){\makebox(0,0)[lb]{\smash{{\SetFigFont{12}{14.4}{\rmdefault}{\mddefault}{\updefault}{\color[rgb]{0,0,0}${\bf\hat{m}}_s$}%
}}}}
\put(2286,-6439){\makebox(0,0)[lb]{\smash{{\SetFigFont{11}{13.2}{\rmdefault}{\mddefault}{\updefault}{\color[rgb]{0,0,0}$\mcA_s$}%
}}}}
\put(2332,-6986){\makebox(0,0)[lb]{\smash{{\SetFigFont{12}{14.4}{\rmdefault}{\mddefault}{\updefault}{\color[rgb]{0,0,0}$B$}%
}}}}
\put(4274,-5114){\makebox(0,0)[lb]{\smash{{\SetFigFont{11}{13.2}{\rmdefault}{\mddefault}{\updefault}{\color[rgb]{0,0,0}$h_t$}%
}}}}
\put(2419,-5524){\makebox(0,0)[lb]{\smash{{\SetFigFont{11}{13.2}{\rmdefault}{\mddefault}{\updefault}{\color[rgb]{0,0,0}$U_n$}%
}}}}
\put(3006,-5246){\makebox(0,0)[lb]{\smash{{\SetFigFont{12}{14.4}{\rmdefault}{\mddefault}{\updefault}{\color[rgb]{0,0,0}${\bf U}$}%
}}}}
\put(1226,-5271){\makebox(0,0)[lb]{\smash{{\SetFigFont{12}{14.4}{\rmdefault}{\mddefault}{\updefault}{\color[rgb]{0,0,0}$z$}%
}}}}
\put(1971,-7555){\makebox(0,0)[lb]{\smash{{\SetFigFont{12}{14.4}{\rmdefault}{\mddefault}{\updefault}{\color[rgb]{0,0,0}$x_1$}%
}}}}
\put(4306,-7555){\makebox(0,0)[lb]{\smash{{\SetFigFont{12}{14.4}{\rmdefault}{\mddefault}{\updefault}{\color[rgb]{0,0,0}$x_2$}%
}}}}
\put(3199,-5891){\makebox(0,0)[lb]{\smash{{\SetFigFont{12}{14.4}{\rmdefault}{\mddefault}{\updefault}{\color[rgb]{0,0,0}$\dot x$}%
}}}}
\put(4829,-5545){\makebox(0,0)[lb]{\smash{{\SetFigFont{12}{14.4}{\rmdefault}{\mddefault}{\updefault}{\color[rgb]{0,0,0}${\bf\hat{m}}_s$}%
}}}}
\put(5016,-5155){\makebox(0,0)[lb]{\smash{{\SetFigFont{12}{14.4}{\rmdefault}{\mddefault}{\updefault}{\color[rgb]{0,0,0}${\bf f}_\gamma$}%
}}}}
\put(1408,-6239){\makebox(0,0)[lb]{\smash{{\SetFigFont{12}{14.4}{\rmdefault}{\mddefault}{\updefault}{\color[rgb]{0,0,0}${\bf f}_\gamma$}%
}}}}
\put(2974,-4929){\makebox(0,0)[lb]{\smash{{\SetFigFont{11}{13.2}{\rmdefault}{\mddefault}{\updefault}{\color[rgb]{0,0,0}${\bf U}_s$}%
}}}}
\put(3764,-5184){\makebox(0,0)[lb]{\smash{{\SetFigFont{12}{14.4}{\rmdefault}{\mddefault}{\updefault}{\color[rgb]{0,0,0}$W$}%
}}}}
\put(1391,-7211){\makebox(0,0)[lb]{\smash{{\SetFigFont{12}{14.4}{\rmdefault}{\mddefault}{\updefault}{\color[rgb]{0,0,0}${\bf U}_C$}%
}}}}
\put(4746,-7556){\makebox(0,0)[lb]{\smash{{\SetFigFont{12}{14.4}{\rmdefault}{\mddefault}{\updefault}{\color[rgb]{0,0,0}$x$}%
}}}}
\put(4913,-7212){\makebox(0,0)[lb]{\smash{{\SetFigFont{12}{14.4}{\rmdefault}{\mddefault}{\updefault}{\color[rgb]{0,0,0}${\bf U}_C$}%
}}}}
\end{picture}%